\documentclass[a4paper]{article}%
\usepackage{amsmath,amssymb,graphicx,epsfig,epic,mathrsfs}
\usepackage{hyperref}
\usepackage[T1]{fontenc}
\usepackage{amsmath}
\usepackage{amsfonts}
\usepackage{amssymb}
\usepackage{graphicx}%
\providecommand{\U}[1]{\protect\rule{.1in}{.1in}}
\newcommand{\field}[1]{\mathbb{#1}}
\newcommand{\rz}{\field{R}}

\newtheorem{theorem}{Theorem}[section]
\newtheorem{lemme}[theorem]{Lemma}

\newtheorem{proposition}[theorem]{Proposition}

\makeatletter

\@addtoreset{equation}{section}
\makeatother
\begin{document}

\title{Time-dependent delta-interactions for 1D Schr\"{o}dinger Hamiltonians.}
\author{T. Hmidi \thanks{IRMAR, UMR-CNRS 6625, Universit\'e Rennes 1, Campus de
Beaulieu, 35042 Rennes Cedex, France, andrea.mantile@univ-rennes1.fr}, A.
Mantile \thanks{IRMAR, UMR-CNRS 6625, Universit\'e Rennes 1, Campus de
Beaulieu, 35042 Rennes Cedex, France, andrea.mantile@univ-rennes1.fr}, F. Nier
\thanks{IRMAR, UMR-CNRS 6625, Universit\'e Rennes 1, Campus de Beaulieu, 35042
Rennes Cedex, France, Francis.Nier@univ-rennes1.fr}}
\date{}
\maketitle

\begin{abstract}
The non autonomous Cauchy problem $i\partial_{t}u=-\partial_{xx}^{2}%
u+\alpha(t) \delta_{0}u$ with $u_{t=0}=u_{0}$ is considered in $L^{2}%
(\mathbb{R})$\,.  The regularity assumptions for $\alpha$ are accurately
analyzed and show that the general results for non autonomous linear
evolution equations in Banach spaces are far from being optimal. In the mean
time, this article shows an unexpected application of paraproduct techniques,
initiated by J.M.~Bony for nonlinear partial differential equations, to a
classical linear problem.

\end{abstract}

\noindent\textit{keywords:} Point interactions, solvable models in Quantum
Mechanics, non autonomous Cauchy problems.

\section{Introduction}

This work is concerned with the dynamics generated by the particular class of
non-autonomous quantum Hamiltonians: $H_{\alpha(t)}=-\frac{d^{2}}{dx^{2}%
}+\alpha(t)\delta$, defining the time dependent delta shaped perurbations of
the 1D Laplacian. Quantum hamiltonians with point interactions  were first
introduced by physicists as a computational tool to study the scattering of
quantum particles with small range forces. Since then, the subject has been
widely developed both in the theoretical framework as well as in the
applications (we refer to \cite{Albeverio} for an extensive presentation).
For real values of the coupling parameter $\alpha$, the rigorous definition
of: $H_{\alpha}=-\frac{d^{2}}{dx^{2}}+\alpha\delta$ arises from the Krein's
theory of selfadjoint extentions. In particular, $H_{\alpha}$ identifies with
the selfadjoint extension of the symmetric operator: $H_{0}=-\frac{d^{2}%
}{dx^{2}}$, $D(H_{0})=C_{0}^{\infty}(\mathbb{R}\backslash\left\{  0\right\}
)$ defined through the boundary conditions%
\begin{equation}
\left\{
\begin{array}
[c]{l}%
\psi^{\prime}(0^{+})-\psi^{\prime}(0^{-})=\alpha\psi(0)\\
\psi(0^{+})-\psi(0^{-})=0
\end{array}
\right.  \label{boundary condition}%
\end{equation}
$\psi(0^{\pm})$ denoting the right and left limit values of $\psi(x)$ as
$x\rightarrow0$ \cite{Albeverio}. Explicitely, one has%
\begin{equation}
D(H_{\alpha})=\left\{  \psi\in H^{2}(\mathbb{R}\backslash\left\{  0\right\}
)\cap H^{1}(\mathbb{R})\,\left\vert \,\psi^{\prime}(0^{+})-\psi^{\prime}%
(0^{-})=\alpha\psi(0)\right.  \right\}
\end{equation}%
\begin{equation}
H_{\alpha}\psi=-\frac{d^{2}}{dx^{2}}\psi\qquad\text{in }\mathbb{R}%
\backslash\left\{  0\right\}.
\end{equation}
When $\alpha(t)$ is assigned as a real valued function of time, the domain
$D(H_{\alpha(t)})$ changes in time with the boundary condition
(\ref{boundary condition}), while the form domain is given by $H^{1}%
(\mathbb{R})$. The quantum evolution associated to the family of operators
$\left\{  H_{\alpha(t)}\right\}  $ is defined by the solutions to the equation%
\begin{equation}
\left\{
\begin{array}
[c]{l}%
i\frac{d}{dt}u=H_{\alpha(t)}u\\
u_{|t=0}=u_{0}\,.%
\end{array}
\right.  \label{eq.Cauchy}%
\end{equation}
The mild solutions are the solutions to the associated integral equation
\begin{equation}
u(t)=e^{it\Delta}u_{{0}}-i\int_{{0}}^{t}e^{i(t-s)\Delta}q(s)\delta_{0}~ds
\label{eq.mildint}%
\end{equation}
with $q(s)=\alpha(s)u(s,0)$\thinspace. The questions are about:

\begin{itemize}
\item the regularity assumptions on $t\to\alpha(t)$ for which
\eqref{eq.Cauchy} defines a unitary strongly continuous dynamical system on
$L^{2}(\mathbb{R})$

\item the meaning of the differential equation \eqref{eq.Cauchy}, according to
the regularity of $t\rightarrow\alpha(t)$\thinspace.
\end{itemize}

General conditions for the solution of this class of problems have been long
time investigated. In the framework of evolution equations in Banach spaces,
Kato was the first who obtained a result for the Cauchy problem%
\begin{equation}
\left\{
\begin{array}
[c]{l}%
\frac{d}{dt}u=A(t)u\\
u_{t=0}=u_{0}%
\end{array}
\right.  , \label{Cauchy}%
\end{equation}
when $t\rightarrow A(t)$ is an unbounded operator valued function \cite{Kat}.
This result, which applies to the quantum dynamical case for $A(t)=-iH(t)$,
requires the strong differentiability of the function $t\rightarrow A(t)$ and
the time independence of the domain $D(A(t))$. Afterwards, a huge literature
was devoted to this problem in the main attempt of relaxing the above conditions
(e.g. in \cite{Simon} and \cite{Reed}; a rather large bibliography and an
extensive presentation of the subject are also given in \cite{Fattorini}). In
particular, the time dependent domain case was explicitely treated in
\cite{Kys} by Kisy\'{n}ski using coercivity assumptions and $C_{loc}^{2}%
$-regularity of $t\rightarrow A(t)$. The regularity conditions in time were
substantially relaxed into a later work of Kato \cite{Kato1} who proves weak
and strong existence results, for the solutions to (\ref{Cauchy}), when
$\left\{  A(t)\right\}  $ forms a stable (a notion defined in \cite{Kato1}%
-Definition~2.1 expressing uniform bounds for the norms of resolvents) family
of generators of contraction semigroups leaving invariant a dense set $Y$ of
the Banach space $X$, and the map $t\rightarrow A(t)$ is norm-continuous in
$\mathcal{L}(Y,X)$\,.

Due to the particular structure, point interaction Hamiltonians allows rather
explicitely energy and resolvent estimates, so that most of the techniques
employed in the analysis of non-autonomous Hamiltonians can be used to deal
with the equation (\ref{eq.Cauchy}), provided that $\alpha(t)$ is regular
enough. At this concern, the Yafaev's works \cite{Yaf1}, \cite{Yaf2} and
\cite{Yaf3} (with M.~Sayapova) on the scattering problems for time dependent
delta interactions in the 3D setting  are to be recalled: There the condition
$\alpha\in C_{loc}^{2}(t_{0},+\infty)$ is used to ensure the existence of a
strongly differentiable time propagator for the quantum evolution. Such a
condition, however, could be considerably relaxed. In our case, for instance,
a first appraoch consist in adapting the strategy of \cite{Kys} by
constructing a family of unitary maps $V_{t,t_{0}}$ such that: $V_{t,t_{0}%
}H_{\alpha(t)}V_{t,t_{0}}^{\ast}$ has a constant domain; then, it is possible
to solve the evolution problem for the deformed operator by using results from
\cite{Kato1}. To fix the idea, let $y_{t}$ be the time dependent vector field
defined by%
\begin{equation}
\left\{
\begin{array}
[c]{l}%
\dot{y}_{t}=g(y_{t},t)\\
y_{t_{0}}=x
\end{array}
\right.  \label{loc_dil}%
\end{equation}
with $g(\cdot,t)\in C^{0}\left(  0,T;\,C_{0}^{\infty}(\mathbb{R})\right)  $,
$\mathrm{supp}~ g(\cdot,t)\subset\left(  -1,1\right)  $, and $g(0,t)=0$ for
each $t$. Under these assumptions, (\ref{loc_dil}) allows an unique solution
depending continuously from time and Cauchy data $\left\{  t_{0},x\right\}  $.
Using the notation: $y_{t}=F(t,t_{0},x)$, one has%
\begin{equation}
\partial_{x}F(t,t_{0},x)=e^{\int_{t_{0}}^{t}\partial_{1}g(y_{s},s)ds}%
>0\qquad\forall x\in\mathbb{R}\,, \label{loc_dil 1}%
\end{equation}
$\partial_{1}g(\cdot,s)$ denoting the derivative w.r.t. the first variable.
This condition allows to consider the map of $x\rightarrow F(t,t_{0},x)$ as a
time-dependent local dilation and one can construct the family of time
dependent unitary transformation associated to it%
\[
\left\{
\begin{array}
[c]{l}%
\medskip\left(  V_{t_{0},t}u\right)  (x)=\left(  \partial_{x}F(t,t_{0}%
,x)\right)  ^{\frac{1}{2}}u(F(t,t_{0},x))\\
V_{t,t_{0}}=V_{t_{0},t}^{-1}%
\end{array}
\right.
\]
Under the action of $V_{t,t_{0}}$, the equation (\ref{eq.Cauchy}) reads as%
\begin{equation}
\left\{
\begin{array}
[c]{l}%
\medskip i\frac{d}{dt}v=V_{t,t_{0}}H_{\alpha(t)}V_{t_{0},t}v-i\left(  \frac
{1}{2}[\partial_{y}g](y,t)+g(y,t)\partial_{y}\right)  v\\
v_{t=0}=u_{0}%
\end{array}
\right.  \label{eq.Cauchy_def}%
\end{equation}
with%
\[
V_{t,t_{0}}H_{\alpha(t)}V_{t_{0},t}=-\partial_{y}b^{2}\partial_{y}%
+a^{2}-\left(  \partial_{y}ab\right)  +b_{0}\alpha(t)\delta
\]%
\[
b(y,t)=e^{\int_{t_{0}}^{t}\partial_{1}g\left(  F(s,t,y),s\right)  \,ds};\quad
b_{0}=b(0,t)
\]%
\[
a(y,t)=\frac{1}{2}\int_{t_{0}}^{t}\partial_{1}^{2}g\left(  F(s,t,y),s\right)
\,e^{\int_{t_{0}}^{t}\partial_{1}g\left(  F(s^{\prime},t,y),s^{\prime}\right)
\,ds^{\prime}}\,ds
\]
and $V_{t,t_{0}}u=v$. Set: $A(t)=iV_{t,t_{0}}H_{\alpha(t)}V_{t_{0},t}+\left(
\frac{1}{2}\partial_{y}g(y,t)+g(y,t)\partial_{y}\right)  $, the domain
$D(A(t))$ is the subspace of $H^{2}(\mathbb{R}\backslash\left\{  0\right\}
)\cap H^{1}(\mathbb{R})$ identified by the boundary condition%
\[
b(0,t)\left[  u^{\prime}(0^{+})-u^{\prime}(0^{-})\right]  =\alpha(t)u(0)
\]
For $\alpha\in W^{1,1}(0,T)$ and $sign(\alpha(t))=const.$, one can determine
(infinitely many) $g(y,t)$ such that: $\frac{1}{k}b(0,t)=\alpha(t)$ for a
fixed constant $k$. With this choice, the operator's domain is constant,
$D(A(t))=Y$. Moreover, under the same conditions on $\alpha$, one can shows
that $A(t)$ defines a stable family of skew-adjoint operators $t$-continuous
in $\mathcal{L}(H^{1},H^{-1})$-operator norm. Thus, one can use the Theorem
5.2 and Remark 5.3 in \cite{Kato1} to get strongly solutions for the related
evolution problem. This is summarized in the following Proposition.

\begin{proposition}
Let $\alpha\in W^{1,1}(0,T)$, $sign(\alpha(t))=const.$ and $u_{0}\in
D(H_{\alpha(0)})$. There exists an unique solution $u_{t}$ of the problem
(\ref{eq.Cauchy}), with: $u_{t}\in D(H_{\alpha(t)})$ for each $t$ and
$u_{t}^{\prime}\in C^{0}(0,T;\,L^{2}(\mathbb{R}))$.
\end{proposition}

In spite of those improved results with already known general tools, our aim
is to prove that they are far from being optimal. An additional structure can
lead to the same conclusions with weaker regularity assumptions. This is the
question that we propose to explore with one dimensional $\delta$-interactions
which allow direct and explicit computations. The main result of this paper is
the following.

\begin{theorem}
\label{th.deptps} \noindent\textbf{$1)$} Assume $\alpha\in H_{loc}^{\frac
{1}{4}}(\mathbb{R})$ , then for any $u_{{0}}\in H^{1}(\mathbb{R})$ the
integral equation (\ref{eq.mildint}) admits a unique solution $u\in
C(\mathbb{R};H^{1}(\mathbb{R}))$ with $i\partial_{t}u-\alpha(t)u(t,0)\delta
_{0}\in C(\mathbb{R};H^{-1}(\mathbb{R}))$ and (\ref{eq.Cauchy}) is weakly
well-posed.\newline\noindent\textbf{$2)$} With the same assumption,
(\ref{eq.mildint}) defines a unitary strongly continuous dynamical system
$U(t,s)$ on $L^{2}(\mathbb{R})$\thinspace.\newline\noindent\textbf{$3)$} If
additionally $\alpha\in H_{loc}^{3/4}(\mathbb{R})$, then for any $u_{0}\in
D(H_{\alpha(0)})$ the solution $u$ of (\ref{eq.mildint}) belongs to the space
$C^{1}(\mathbb{R};L^{2}(\mathbb{R})),$ with $u(t)\in D(H_{\alpha(t)})$ for
every $t\in\mathbb{R}.$
\end{theorem}

\begin{remark}
The problem of defining the quantum evolution for 1D time dependent delta
interactions has also been considered in a nonlinear setting \cite{Adami}
where $\alpha$ is assigned as a function of the particle's state, in our
notation: $\alpha(t)=\gamma\left\vert u_{t}(0)\right\vert ^{2\sigma}$, with
$\gamma\in\mathbb{R}$, $\sigma\in\mathbb{R}_{+}$. In this framework, the
authors prove that solutions to the nonlinear evolution problem exist locally
in time for $u_{0}\in H^{\rho}$ with $\rho>\frac{1}{2}$ and: $U(t,s)u_{0}%
(0)\in H_{loc}^{\frac{\rho}{2}+\frac{1}{4}}$ as a function of time. This
corresponds to the condition: $\alpha\in H_{loc}^{\nu}$ with $\nu>\frac{1}{2}%
$. Although a linear problem is considered here, the result of
Theorem~\ref{th.deptps}
improves significantly the regularity condition: $\alpha\in
H^{1/4}_{loc}(\rz)$ is weaker than a continuity assumption.
\end{remark}

In what follows, $D_{x}$ denotes $\frac{1}{i}\partial_{x}=\mathcal{F}%
^{-1}\circ(\xi\times)\circ\mathcal{F}$, where $\mathcal{F}$ is the Fourier
transform in position (or in time) normalized according to%
\[
\mathcal{F}\varphi(\xi)=\int_{\mathbb{R}}e^{-i\xi x}\varphi(x)~dx\,.
\]
The Sobolev spaces are denoted by $H^{s}(\mathbb{R})$, $s\in\mathbb{R}$, their
local version by $H_{loc}^{s}(\mathbb{R})$. The notation $u\in H^{s+0}%
(\mathbb{R})$ means that there exists $\varepsilon>0$ such that $u\in
H^{s+\varepsilon}(\mathbb{R})$ (inductive limit) and its local version $u\in
H_{loc}^{s+0}(\mathbb{R})$ allows $\varepsilon_{R}>0$ to depend on $R>0$ while
considering the interval $\left[  -R,R\right]  $. More generally the Besov
spaces are defined through dyadic decomposition: For $(p,r)\in\lbrack
1,+\infty]^{2}$ and $s\in\mathbb{R},$ the space $B_{p,r}^{s}$ is the set of
tempered distributions $u$ such that%
\[
\Vert u\Vert_{B_{p,r}^{s}}:=\Big(2^{qs}\Vert\Delta_{q}u\Vert_{L^{p}%
}\Big)_{\ell^{r}}<+\infty\,,
\]
where $\Delta_{q}=\varphi(2^{-q}D)$, $q\in\mathbb{N}$, is a cut-off in the
Fourier variable supported in $C^{-1}2^{q}\leq|\xi|\leq C2^{q}$\thinspace.
Details are given in Appendix~\ref{se.law}. Finally, the notation '$\lesssim
$', appearing in many of the following proofs, denotes the inequality: '$\leq
C$', being $C$ a suitable positive constant.

\section{Proof of Theorem~\ref{th.deptps}}

\label{se.proof}

This theorem is a consequence of simple remarks, explicit calculations and
standard applications of paraproduct estimates in 1D Sobolev spaces. Let us
start with some elementary rewriting of the Cauchy problem \eqref{eq.Cauchy}.

\subsection{Preliminary remarks}

\label{se.prel}

\begin{itemize}
\item First of all equation \eqref{eq.Cauchy} or its integral version
\eqref{eq.mildint} are local problems in time so that $t_{0}=0$, $t\in[-T,T]$
for some $T>0$ and even $\mathrm{supp}~ \alpha\subset[-T/2,T/2]$ can be
assumed after replacing $\alpha$ with $\alpha_{T}(s)=\alpha(s)\chi(\frac{s}%
{T})$ for some fixed $\chi\in\mathscr{C}^{\infty}_{0}((-1/2,1/2))$ and
$\chi\equiv1$ near $s=0$. The dependence of $H^{s}$-norms of $\alpha_{T}$ with
respect to $T$ will be discussed when necessary.

\item The equation \eqref{eq.Cauchy} or its integral version
\eqref{eq.mildint} makes sense in $\mathcal{S}^{\prime}(\mathbb{R}_{x})$ as
soon as $u(t,0)$ is well defined for almost all $t\in\lbrack-T,T]$ and
$q(t)=\alpha_{T}(t)u(t,0)$ is locally integrable. Then it can be written after
applying the Fourier transform as a local problem in $\xi\in\mathbb{R}$
\begin{align}
&  \left\{
\begin{array}
[c]{l}%
i\partial_{t}\widehat{u}(t,\xi)=|\xi|^{2}\widehat{u}(t,\xi)+q(t)\\
\widehat{u}(0,\xi)=\widehat{u}_{0}(\xi)
\end{array}
\right.  \label{eq.four1}\\
\text{with\,\, } &  q(t)=\alpha_{T}(t)u(t,0)=\alpha_{T}(t)\int_{\mathbb{R}}\widehat
{u}(t,\xi)~d\xi\,.
\end{align}
This is equivalent to the integral form
\begin{align}
&  \widehat{u}(t,\xi)=e^{-it|\xi|^{2}}\widehat{u}_{0}(\xi)-i\int_{0}^{t}%
e^{-i(t-s)|\xi|^{2}}q(s)~ds\label{eq.convol}\\
\text{with\,\,  } &  q(t)=q_{0}(t)-i\alpha_{T}(t)\int_{0}^{t}\int_{\mathbb{R}%
}e^{-i(t-s)|\xi|^{2}}q(s)~dsd\xi\,
\end{align}
by setting $q_{0}(t)=\alpha_{T}(t)[e^{it\Delta}u_{0}](0)$\thinspace. The
assumption $u_{0}\in H^{s}(\mathbb{R})$, $s>1/2$, (resp. $u_{0}\in
L^{1}(\mathbb{R})$) ensures that $[e^{it\Delta}u_{0}](0)\in C^{0}([-T,T])$
(resp. $t^{1/2}[e^{it\Delta}u_{0}](0)\in C^{0}([-T,T])$). Such an assumption
as well as looking for $u(t)\in H^{1}(\mathbb{R})$ ensures that the quantities
$q_{0}(t)$ and $q(t)$ make sense for almost all $t\in\lbrack-T,T]$.

\item With the support assumption $\mathrm{supp}~\alpha_{T}\subset
\lbrack-T/2,T/2]$, the convolution equation (\ref{eq.convol}) can be written
\begin{align}
q(t) &  =q_{0}(t)-i\alpha_{T}(t)\int_{0}^{t}\int_{\mathbb{R}}1_{[-T,T]}%
(t-s)e^{-i(t-s)|\xi|^{2}}1_{[-T,T]}(s)q(s)~dsd\xi\quad\text{in}\quad
\mathcal{D}^{\prime}(\mathbb{R})\nonumber\\
&  :=q_{0}(t)-i\alpha_{T}(t)\mathcal{L}q(t):=q_{0}(t)+\mathcal{L}_{\alpha}%
q(t)\,.\label{eq.convolsupp}%
\end{align}

\item Once $q$ is known after solving (\ref{eq.convolsupp}), equation
(\ref{eq.convol}) with $t\in\rz$ reads simply
\begin{equation}
\widehat{u}(t,\xi)=e^{-it|\xi|^{2}}\widehat{u}_{0}(\xi)-ie^{-it|\xi|^{2}}%
\mathcal{F}\left[  q1_{[0,t]}\right]  (-|\xi|^{2})\,.\label{eq.mildfour2}%
\end{equation}

\item When $u_{0}$ and $q$ are regular enough the time-derivative of the
quantity (\ref{eq.convol}) gives
\[
i\partial_{t}(\partial_{t}\widehat{u})(t,\xi)=|\xi|^{2}\partial_{t}\widehat{u}%
(t,\xi)+q^{\prime}(t).
\]
By Duhamel formula, this implies
\[
\partial_{t}\widehat{u}(t,\xi)=e^{-it|\xi|^{2}}\partial_{t}\widehat{u}(0,\xi
)-i\int_{0}^{t}e^{-i(t-s)|\xi|^{2}}q^{\prime}(s)ds\,,
\]
while (\ref{eq.four1}) says for $t=0$
\[
\partial_{t}\widehat{u}(0,\xi)=-i|\xi|^{2}\widehat{u}_{0}(\xi)-iq(0)\,.
\]
Therefore we obtain for $t\in\mathbb{R}$%
\begin{equation}\label{eq.timeder2}
i\partial_{t}\widehat{u}(t,\xi)=e^{-it|\xi|^{2}}\left[  |\xi|^{2}\widehat{u}_{0}%
(\xi)+q(0)\right]  +e^{-it|\xi|^{2}}\mathcal{F}\left[  (\partial
_{s}q)1_{[0,t]}\right]  (-|\xi|^{2})\,.
\end{equation}

\end{itemize}


\subsection{Reduced scalar equation for $q$}

\label{se.redq} Let us now study the equation (\ref{eq.convolsupp}) written:
\[
q=q_{0}+\mathcal{L}_{\alpha}q
\]
with
\[
\mathcal{L}_{\alpha}q:=-i\alpha_{T}(t)\mathcal{L}q=-i\alpha_{T}(t)\int_{0}^{t}%
\int_{\mathbb{R}}1_{[-T,T]}(t-s)e^{-i(t-s)|\xi|^{2}}1_{[-T,T]}(s)q(s)~dsd\xi
\,.
\]
Solving this fixed point equation relies on the next result.

\begin{proposition}
\label{smoothing} The estimate
\[
\|\mathcal{L} q\|_{H^{s}}\lesssim T^{\frac12} \|1_{[-1,1]}(\mathrm{D}%
_{t})q\|_{L^{2}}+T^{\frac{1}{2}-{\theta}} \Big(
\|q\,1_{[0,T]} \|_{H^{s-{\theta}}}+\|q\,1_{[-T,0]}\|_{H^{s-{\theta}}}\Big)
\]
holds for every $s\in\mathbb{R}$ and $\theta\in[0,\frac12]$\,.
\end{proposition}

\noindent\textbf{Proof: }Owing to ${\int_{\mathbb{R}}e^{\pm i\lambda|\xi|^{2}%
}~\frac{d\xi}{2\pi}=\frac{e^{\pm i\frac{\pi}{4}}}{\sqrt{4\pi\lambda}}}$ for
$\lambda>0,$ $\mathcal{L}q$ writes as%
\[
\frac{1}{\sqrt{\pi}}\mathcal{L}q(t)=e^{-i\frac{\pi}{4}}\int_{\mathbb{R}%
}1_{\left[  0,T\right]  }(s)q(s)\frac{1_{\left[  0,T\right]  }(t-s)}{\left(
t-s\right)  ^{\frac{1}{2}}}ds+e^{i\frac{\pi}{4}}\int_{\mathbb{R}}1_{\left[
-T,0\right]  }(s)q(s)\frac{1_{\left[  -T,0\right]  }(t-s)}{\left(  s-t\right)
^{\frac{1}{2}}}ds
\]
Passing to the Fourier transform, we get%
\[
\frac{1}{\sqrt{\pi}}\widehat{\mathcal{L}q}(\tau)=e^{-i\frac{\pi}{4}}
\left(\int
_{0}^{T}\frac{e^{-it\tau}}{\sqrt{t}}~dt\right)\,\mathcal{F}(1_{[0,T]}q)(\tau
)+e^{i\frac{\pi}{4}}
\left(
\int_{0}^{T}\frac{e^{it\tau}}{\sqrt{t}}~dt
\right)\,\mathcal{F}%
(1_{[-T,0]}q)(\tau)
\]
One easily checks
\[
\left\vert \int_{0}^{T}\frac{e^{\pm it\tau}}{\sqrt{t}}dt\right\vert \leq
2\sqrt{T}\quad\text{and}\quad\left\vert \int_{0}^{T}\frac{e^{\pm it\tau}%
}{\sqrt{t}}dt\right\vert \lesssim|\tau|^{-\frac{1}{2}}.
\]
This yields for every $\theta\in\lbrack0,\frac12],\tau\in\mathbb{R}$
\[
\left\vert \int_{0}^{T}\frac{e^{\pm it\tau}}{\sqrt{t}}dt\right\vert \lesssim
T^{\frac{1}{2}}1_{[-1,1]}(\tau)+T^{\frac{1}{2}-\theta}|\tau|^{-
{\theta}}1_{\left\{  \mathbb{R}\setminus\lbrack-1,1]\right\}  }(\tau).
\]
Thus we get for every $s\in\mathbb{R},\theta\in\lbrack0,\frac12],$
\[
\Vert\mathcal{L}q\Vert_{H^{s}}\lesssim T^{\frac{1}{2}}\Vert1_{[-1,1]}%
(D_{t})q\Vert_{L^{2}}+T^{\frac{1}{2}-\theta}\Big(\Vert1_{[0,T]}%
q\Vert_{H^{s-{\theta}}}+\Vert1_{[-T,0]}q\Vert_{H^{s-{\theta}}%
}\Big).
\]
{\hfill$\square$\newline}

\begin{proposition}
\label{sol1}

\begin{enumerate}
\item Let $u_{0}\in H^{s}(\mathbb{R}_{x})$ with $s>1/2$ and let $\alpha\in
H_{loc}^{\frac{1}{4}}(\mathbb{R}_{t})$\thinspace. Then the equation
(\ref{eq.convolsupp}) has a unique solution $q\in H_{loc}^{\frac{1}{4}%
}(\mathbb{R}_{t})$. Moreover, for a fixed $u_{0}\in H^{s}(\mathbb{R}_{x})$
with $s>1/2$, the map $\alpha\mapsto q$ is locally Lipschitzian from
$H^{\frac{1}{4}}(\mathbb{R}_{t})$ to $H^{\frac{1}{4}}(\mathbb{R}_{t}%
)$\thinspace.

\item Let $u_{0}\in H^{1}(\mathbb{R}_{x})$ and let $\alpha\in H_{loc}%
^{\frac{3}{4}}(\mathbb{R}_{t})$\thinspace. Then the equation
(\ref{eq.convolsupp}) has a unique solution $q\in H_{loc}^{\frac{3}{4}%
}(\mathbb{R}_{t})$. Moreover, for a fixed $u_{0}\in H^{1}(\mathbb{R}_{x})$ the
map $\alpha\mapsto q$ is locally Lipschitzian from $H^{\frac{3}{4}}%
(\mathbb{R}_{t})$ to $H^{\frac{3}{4}}(\mathbb{R}_{t})$.

\item Let $u_{0}\in H^{1+2\varepsilon}(\mathbb{R}_{x})$ and let $\alpha\in
H_{loc}^{\frac{3}{4}+\varepsilon}(\mathbb{R}_{t})$\thinspace, for some
$\varepsilon>0$. Then the equation (\ref{eq.convolsupp}) has a unique solution
$q\in H_{loc}^{\frac{3}{4}+\varepsilon}(\mathbb{R}_{t}).$
\end{enumerate}
\end{proposition}



\noindent\textbf{Proof: \ } \textbf{1)} Let us first prove that $q_{0}\in
H^{\frac{1}{4}}$ when $u_{0}\in H^{\frac{1}{2}+\varepsilon}$\thinspace. Write
first
\begin{align*}
(e^{it\Delta}u_{0})(0) &  =\int_{-1}^{1}e^{-it|\xi|^{2}}\widehat{u_{0}}%
(\xi)~\frac{d\xi}{2\pi}+\int_{1}^{+\infty}e^{-it\tau}\left[  \widehat{u_{0}%
}(\sqrt{\tau})+\widehat{u_{0}}(-\sqrt{\tau})\right]  ~\frac{d\tau}{2\pi
\sqrt{\tau}}\\
&  =I(t)+II(t)\,.
\end{align*}
The first term $I(t)$ defines a $C^{\infty}$ function with%
\[
\Vert I\Vert_{B_{\infty,\infty}^{s}}\lesssim\Vert u_{0}\Vert_{L^{2}}\,,
\]
for every $s\in\mathbb{R}$\thinspace. On the other hand, the Fourier transform
of $II$ equals
\[
\widehat{II}(-\tau)=1_{[1,\infty\lbrack}(\tau)\left(  \widehat{u_{0}}%
(\sqrt{\tau})+\widehat{u_{0}}(-\sqrt{\tau})\right)  ~\frac{1}{\sqrt{\tau}}\,.
\]
The Sobolev regularity of the second term, $II$, is given by:
\begin{align}
\Vert II\Vert_{H^{\nu}}^{2} &  \lesssim\int_{1}^{+\infty}\tau^{2\nu}\left\vert
\frac{\widehat{u_{0}}(\sqrt{\tau})}{\sqrt{\tau}}\right\vert ^{2}%
~d\tau\nonumber\label{sob1}\\
&  \lesssim\left\Vert u_{0}\right\Vert _{H^{2\nu-1/2}}^{2}\,,
\end{align}
for any $\nu\in\mathbb{R}$\thinspace. Now, write
\[
q_{0}(t)=\alpha_{T}(t)I(t)+\alpha_{T}(t)II(t).
\]
Lemma \ref{paraproduct}-b) applied to the first term, implies
\begin{align*}
\Vert\alpha_{T}I\Vert_{H^{\frac{1}{4}}} &  \lesssim\Vert\alpha_{T}%
\Vert_{H^{\frac{1}{4}}}\Vert I\Vert_{B_{\infty,\infty}^{s+\varepsilon}}\\
&  \lesssim\Vert\alpha_{T}\Vert_{H^{\frac{1}{4}}}\Vert u_{0}\Vert_{L^{2}}.
\end{align*}
For the second term, use Lemma \ref{paraproduct}-a), Sobolev embeddings and
(\ref{sob1})
\begin{align*}
\Vert\alpha_{T}II\Vert_{H^{\frac{1}{4}}} &  \lesssim\Vert\alpha_{T}%
\Vert_{H^{\frac{1}{4}}}\Vert II\Vert_{B_{2,\infty}^{\frac{1}{2}}\cap
L^{\infty}}\\
&  \lesssim\Vert\alpha_{T}\Vert_{H^{\frac{1}{4}}}\Vert II\Vert_{H^{\frac{1}%
{2}+\frac{\varepsilon}{2}}}\\
&  \lesssim\Vert\alpha_{T}\Vert_{H^{\frac{1}{4}}}\Vert u_{0}\Vert_{H^{\frac
{1}{2}+\varepsilon}}.
\end{align*}
By combining these estimates, we get
\[
\Vert q_{0}\Vert_{H^{\frac{1}{4}}}\lesssim\Vert\alpha_{T}\Vert_{H^{\frac{1}%
{4}}}\Vert u_{0}\Vert_{H^{\frac{1}{2}+\varepsilon}}\,.
\]
It remains to estimate $\alpha_{T}$. Let $\widetilde{\chi}\in\mathcal{D}%
(\mathbb{R})$ with $\widetilde{\chi}\equiv1$ in $[-1,1]$ and set
$\widetilde{\alpha}(t)=\widetilde{\chi}(t)\alpha(t)$\thinspace. By using again
Lemma \ref{paraproduct}-a), we get for $0\leq T\leq1$
\[
\Vert\alpha_{T}\Vert_{H^{\frac{1}{4}}}\lesssim\Vert\widetilde{\alpha}%
\Vert_{H^{\frac{1}{4}}}\big\|\chi(T^{-1}\cdot)\big\|_{B_{2,\infty}^{\frac
{1}{2}}\cap L^{\infty}}.
\]
A change of variable in the Fourier transform $F\left[  \chi(T^{-1}.)\right]
(\tau)=T\widehat{\chi}(T\tau)$ leads to
\begin{equation}
\left\Vert \chi(T^{-1}.)\right\Vert _{H^{\mu}}\leq T^{\frac{1}{2}-\mu
}\left\Vert \chi\right\Vert _{H^{\mu}}\quad\text{and}\quad\left\Vert
\chi(T^{-1}.)\right\Vert _{B_{2,\infty}^{\mu}}\lesssim T^{\frac{1}{2}-\mu
}\left\Vert \chi\right\Vert _{B_{2,\infty}^{\mu}}\,,\label{eq.dilatfou}%
\end{equation}
for $\mu\geq0$ and $T\leq1$. Hence we get
\[
\Vert\alpha_{T}\Vert_{H^{\frac{1}{4}}}\lesssim\Vert\widetilde{\alpha}%
\Vert_{H^{\frac{1}{4}}}.
\]
and
\begin{equation}
\Vert q_{0}\Vert_{H^{\frac{1}{4}}}\lesssim\Vert\widetilde{\alpha}%
\Vert_{H^{\frac{1}{4}}}\Vert u_{0}\Vert_{H^{\frac{1}{2}+\varepsilon}%
}\label{initial}%
\end{equation}
In order to estimate the the operator $\mathcal{L}$, use Lemma
\ref{paraproduct}-a)
\begin{align*}
\Vert\mathcal{L}_{\alpha}q\Vert_{H^{\frac{1}{4}}} &  \lesssim\Vert\alpha
_{T}\Vert_{H^{\frac{1}{4}}}\left\Vert \mathcal{L}q\right\Vert _{B_{2,\infty
}^{\frac{1}{2}}\cap L^{\infty}}\\
&  \lesssim\Vert\widetilde{\alpha}\Vert_{H^{\frac{1}{4}}}\left\Vert
\mathcal{L}q\right\Vert _{H^{\frac{1}{2}+\varepsilon}}\,,
\end{align*}
while Proposition \ref{smoothing} says
\[
\left\Vert \mathcal{L}q\right\Vert _{H^{\frac{1}{2}+\varepsilon}}\lesssim
T^{\frac{1}{2}}\Vert q\Vert_{L^{2}}+T^{\frac{1}{4}-\varepsilon}\Big(\Vert
1_{[0,T]}q\Vert_{H^{\frac{1}{4}}}+\Vert1_{[-T,0]}q\Vert_{H^{\frac{1}{4}}%
}\Big)\,.
\]
Hence we get for $0\leq T\leq1$ and by Lemma \ref{paraproduct}-a)
\[
\left\Vert \mathcal{L}q\right\Vert _{H^{\frac{1}{2}+\varepsilon}}\lesssim
T^{\frac{1}{4}-\varepsilon}\Vert q\Vert_{H^{\frac{1}{4}}}%
\]
This yields
\begin{equation}
\Vert\mathcal{L}_{\alpha}q\Vert_{H^{\frac{1}{4}}}\lesssim\Vert\widetilde
{\alpha}\Vert_{H^{\frac{1}{4}}}T^{\frac{1}{4}-\varepsilon}\left\Vert
q\right\Vert _{H^{\frac{1}{4}}}.\label{contract}%
\end{equation}
This proves that $\mathcal{L}$ is a contracting map in $H^{\frac{1}{4}}$ for
sufficiently small time $T$. The time $T$ depends only on $\Vert
\widetilde{\alpha}\Vert_{H^{\frac{1}{4}}}$ and then we can construct globally
a unique solution $q\in H_{loc}^{\frac{1}{4}}(\mathbb{R})$ for the linear
problem (\ref{eq.convolsupp}).\newline It remains to prove the continuity
dependence of $q$ with respect to $\alpha$. Let $\alpha,\bar{\alpha}\in
H_{loc}^{\frac{1}{4}}$ and $q,\bar{q}$ the corresponding solutions then we
have
\[
q(t)-\bar{q}(t)=\alpha_{T}(t)\mathcal{L}q(t)-\bar{\alpha}_{T}(t)\mathcal{L}%
\bar{q}(t),\quad\text{with}\quad\bar{\alpha}_{T}(t)=\bar{\alpha}(t)\chi(t/T).
\]
Since $\mathcal{L}$ is linear on $q$ then
\begin{align*}
q(t)-\bar{q}(t) &  =\alpha_{T}(t)\mathcal{L}(q-\bar{q})(t)+(\alpha_{T}%
-\bar{\alpha}_{T})(t)\mathcal{L}\bar{q}(t)\\
&  =\mathcal{L}_{\alpha}(q-\bar{q})(t)+\mathcal{L}_{\alpha-\bar{\alpha}}%
\bar{q}(t).
\end{align*}
To estimate the terms of the r.h.s we use (\ref{contract})
\begin{align*}
\Vert\mathcal{L}_{\alpha}(q-\bar{q})\Vert_{H^{\frac{1}{4}}} &  \lesssim
\Vert\widetilde{\alpha}\Vert_{H^{\frac{1}{4}}}T^{\frac{1}{4}-\varepsilon
}\left\Vert q-\bar{q}\right\Vert _{H^{\frac{1}{4}}},\\
\Vert\mathcal{L}_{\alpha-\bar{\alpha}}\bar{q}\Vert_{H^{\frac{1}{4}}} &
\lesssim\big\|\widetilde{\alpha}-\widetilde{\bar{\alpha}}\big\|_{H^{\frac
{1}{4}}}T^{\frac{1}{4}-\varepsilon}\left\Vert \bar{q}\right\Vert _{H^{\frac
{1}{4}}}.
\end{align*}
With the choice of $T$ done above we get
\[
\Vert q-\bar{q}\Vert_{H^{\frac{1}{4}}}\lesssim\big\|\widetilde{\alpha
}-\widetilde{\bar{\alpha}}\big\|_{H^{\frac{1}{4}}}\Vert\bar{q}\Vert
_{H^{\frac{1}{4}}}.
\]
This achieves the proof of the continuity.

\textbf{2)} Write again $q_{0}(t)=\alpha_{T}(t)I(t)+\alpha_{T}(t)II(t)$. Lemma
\ref{paraproduct}-b) implies
\begin{align*}
\Vert\alpha_{T}I\Vert_{H^{\frac{3}{4}}} &  \lesssim\Vert\alpha_{T}%
\Vert_{H^{\frac{3}{4}}}\Vert I\Vert_{B_{\infty,\infty}^{\frac{3}%
{4}+\varepsilon}}\\
&  \lesssim\Vert\alpha_{T}\Vert_{H^{\frac{3}{4}}}\Vert u_{0}\Vert_{L^{2}}\,.
\end{align*}
Since $H^{\frac{3}{4}}$ is an algebra the inequality
\begin{align*}
\Vert\alpha_{T}\Vert_{H^{\frac{3}{4}}} &  \lesssim\Vert\chi(T^{-1}\cdot
)\Vert_{H^{\frac{3}{4}}}\Vert\widetilde{\alpha}\Vert_{H^{\frac{3}{4}}}\\
&  \lesssim T^{-\frac{1}{4}}\Vert\widetilde{\alpha}\Vert_{H^{\frac{3}{4}}}%
\end{align*}
holds for $T\in\lbrack0,1]$, owing to (\ref{eq.dilatfou}).\newline It follows
\[
\Vert\alpha_{T}I\Vert_{H^{\frac{3}{4}}}\lesssim T^{-\frac{1}{4}}%
\Vert\widetilde{\alpha}\Vert_{H^{\frac{3}{4}}}\Vert u_{0}\Vert_{L^{2}}.
\]
The second term is estimated with (\ref{sob1}):
\begin{align*}
\Vert\alpha_{T}II\Vert_{H^{\frac{3}{4}}} &  \lesssim\Vert\alpha_{T}%
\Vert_{H^{\frac{3}{4}}}\Vert II\Vert_{H^{\frac{3}{4}}}\\
&  \lesssim T^{-\frac{1}{4}}\Vert\widetilde{\alpha}\Vert_{H^{\frac{3}{4}}%
}\Vert u_{0}\Vert_{H^{1}}.
\end{align*}
Finally we get for $T\in\lbrack0,1]$
\[
\Vert q_{0}\Vert_{H^{\frac{3}{4}}}\lesssim T^{-\frac{1}{4}}\Vert
\widetilde{\alpha}\Vert_{H^{\frac{3}{4}}}\Vert u_{0}\Vert_{H^{1}}.
\]
Using Lemma \ref{paraproduct}-a)-d), Sobolev embeddings and Proposition
\ref{smoothing} (with $\theta=\frac{5}{12}$ and $\theta=\frac{1}{6}$), gives
\begin{align}
\Vert\mathcal{L}_{\alpha}q\Vert_{H^{\frac{3}{4}}} &  \lesssim\Vert\alpha
_{T}\Vert_{L^{\infty}}\Vert\mathcal{L}q\Vert_{H^{\frac{3}{4}}}+\Vert\alpha
_{T}\Vert_{H^{\frac{3}{4}}}\Vert\mathcal{L}q\Vert_{L^{\infty}}%
\nonumber\label{cont1}\\
&  \lesssim\Vert\widetilde{\alpha}\Vert_{L^{\infty}}T^{\frac{1}{12}}\big(\Vert
q1_{[0,T]}\Vert_{H^{\frac{1}{3}}}+\Vert q1_{[-T,0]}\Vert_{H^{\frac{1}{3}}%
}\big)+T^{-\frac{1}{4}}\Vert\widetilde{\alpha}\Vert_{H^{\frac{3}{4}}}%
\Vert\mathcal{L}q\Vert_{H^{\frac{1}{2}+\varepsilon}}\nonumber\\
&  \lesssim\Vert\widetilde{\alpha}\Vert_{H^{\frac{3}{4}}}T^{\frac{1}{12}}\Vert
q\Vert_{H^{\frac{1}{3}}}+T^{-\frac{1}{4}}\Vert\widetilde{\alpha}%
\Vert_{H^{\frac{3}{4}}}T^{\frac{1}{3}}\Vert q\Vert_{H^{\frac{1}{3}%
+\varepsilon}}\,.
\end{align}
Thus we get for $0\leq T\leq1$,
\[
\Vert\mathcal{L}_{\alpha}q\Vert_{H^{\frac{3}{4}}}\lesssim\Vert\widetilde
{\alpha}\Vert_{H^{\frac{3}{4}}}T^{\frac{1}{12}}\Vert q\Vert_{H^{\frac{3}{4}}}.
\]
This proves that $\mathcal{L}$ is a contracting map in $H^{\frac{3}{4}}$ for
sufficiently small time $T$. The time $T$ depends only on $\Vert
\widetilde{\alpha}\Vert_{H^{\frac{3}{4}}}$ and then we can construct globally
a unique solution $q\in H_{\mathrm{loc}}^{\frac{3}{4}}(\mathbb{R})$ for the
linear problem.

For the locally Lipschitz dependence with respect to $\alpha$, the proof is
left to the reader: it can be done easily done like for the case $\alpha\in
H^{\frac14}$.

\textbf{3)} Like in the proof of the second point $2)$ we get
\begin{align*}
\Vert q_{0}\Vert_{H^{\frac{3}{4}+\varepsilon}} &  \lesssim\Vert\alpha_{T}%
\Vert_{H^{\frac{3}{4}+\varepsilon}}\big(\Vert u_{0}\Vert_{L^{2}}+\Vert
II\Vert_{H^{\frac{3}{4}+\varepsilon}}\big)\\
&  \lesssim T^{-\frac{1}{4}-\varepsilon}\Vert u_{0}\Vert_{H^{1+2\varepsilon}}.
\end{align*}
Reproducing the same computation as (\ref{cont1}) leads to
\begin{align*}
\Vert\mathcal{L}_{\alpha}q\Vert_{H^{\frac{3}{4}+\varepsilon}} &  \lesssim
\Vert\alpha_{T}\Vert_{L^{\infty}}\Vert\mathcal{L}q\Vert_{H^{\frac{3}%
{4}+\varepsilon}}+\Vert\alpha_{T}\Vert_{H^{\frac{3}{4}+\varepsilon}}%
\Vert\mathcal{L}q\Vert_{L^{\infty}}\\
&  \lesssim\Vert\widetilde{\alpha}\Vert_{L^{\infty}}T^{\frac{1}{12}}\big(\Vert
q1_{[0,T]}\Vert_{H^{\frac{1}{3}+\varepsilon}}+\Vert q1_{[-T,0]}\Vert
_{H^{\frac{1}{3}+\varepsilon}}\big)+T^{-\frac{1}{4}-\varepsilon}%
\Vert\widetilde{\alpha}\Vert_{H^{\frac{3}{4}+\varepsilon}}\Vert\mathcal{L}%
q\Vert_{H^{\frac{1}{2}+\varepsilon}}\\
&  \lesssim\Vert\widetilde{\alpha}\Vert_{L^{\infty}}T^{\frac{1}{12}}\Vert
q\Vert_{H^{\frac{1}{3}+\varepsilon}}+T^{-\frac{1}{4}-\varepsilon}%
\Vert\widetilde{\alpha}\Vert_{H^{\frac{3}{4}+\varepsilon}}T^{\frac{1}%
{3}+\varepsilon}\Vert q\Vert_{H^{\frac{1}{3}+2\varepsilon}}\\
&  \lesssim T^{\frac{1}{12}}\Vert\widetilde{\alpha}\Vert_{H^{\frac{3}%
{4}+\varepsilon}}\Vert q\Vert_{H^{\frac{3}{4}+\varepsilon}}.
\end{align*}
With the fixed point argument we can conclude the proof. {\hfill$\square
$\newline}

\subsection{Regularity of $u$}

\label{se.regu} We start with the following result.

\begin{lemme}
\label{le.chgvar} For $s\in\mathbb{R}$, let $H_{T}^{s}$ be the closed subset
of $H^{s}(\mathbb{R}_{t})$%
\[
H_{T}^{s}=\left\{  u\in H^{s}(\mathbb{R}_{t}),\quad\mathrm{supp}%
~u\subset\lbrack-T,T]\right\}  \,,
\]
endowed with the norm $\left\Vert ~\right\Vert _{H^{s}}$. For any $T>0$ and
any $s\in\mathbb{R}$, there is a constant $C_{T,s}$ such that
\[
\forall f\in H_{T}^{\frac{2s-1}{4}}\,,\quad\left\Vert \mathcal{F}^{-1}\left[
\mathcal{F}f(-|\xi|^{2})\right]  \right\Vert _{H^{s}}\leq C_{T,s}\left\Vert
f\right\Vert _{H^{\frac{2s-1}{4}}}.
\]

\end{lemme}

\noindent\textbf{Proof: \ }{} It suffices to compute
\begin{align*}
\int_{\mathbb{R}}(1+\left\vert \xi\right\vert ^{2})^{s}\left\vert \widehat
{f}(-|\xi|^{2})\right\vert ^{2}~d\xi &  =\int_{0}^{+\infty}\frac{(1+\tau)^{s}%
}{2\tau^{1/2}}\left\vert \widehat{f}(-\tau)\right\vert ^{2}~d\tau\\
&  \leq\max_{\tau\in\lbrack0,1]}\left\vert \widehat{f}(-\tau)\right\vert ^{2}%
+\int_{1}^{\infty}\left(  1+\tau\right)  ^{s-1/2}\left\vert \widehat{f}%
(-\tau)\right\vert ^{2}~d\tau\\
&  \leq\max_{\tau\in\lbrack0,1]}\left\vert \widehat{f}(-\tau)\right\vert
^{2}+\left\Vert f\right\Vert _{H^{\frac{s}{2}-\frac{1}{4}}}^{2}\,,
\end{align*}
where $\widehat{f}(\tau)=\langle e^{i\tau x}\widetilde{\chi}(x)\,,\,f\rangle$
with $\widetilde{\chi}\in\mathcal{D}(\mathbb{R})$ with value $1$ in $[-T,T].$
By duality we have for $\nu\in\mathbb{R}$
\begin{align*}
\sup_{0\leq\tau\leq1}|\widehat{f}(\tau)| &  \leq\Vert f\Vert_{H^{\nu}}%
\sup_{0\leq\tau\leq1}\Vert e^{i\tau\cdot}\widetilde{\chi}(\cdot)\Vert
_{H^{-\nu}}\\
&  \leq C_{T,\nu}^{1}\Vert f\Vert_{H^{\nu}}\,.
\end{align*}
{\hfill$\square$\newline}{} The main result of this section is the following.

\begin{proposition}
\label{propeq}

\begin{enumerate}
\item Let $u_{0}\in H^{1}(\mathbb{R}_{x}),\alpha\in H_{loc}^{\frac{1}{4}%
}(\mathbb{R}_{t}),$ then the equation (\ref{eq.mildint}) has a unique solution
$u\in C(\mathbb{R};H^{1}(\mathbb{R}))$\thinspace.

\item Let $u_{0}\in D(H_{\alpha(0)})$, $\alpha\in H_{loc}^{\frac{3}{4}%
}(\mathbb{R}_{t}),$ then the equation (\ref{eq.mildint}) has a unique solution
$u$ belonging to the space $C^{1}(\mathbb{R};L^{2}(\mathbb{R}))$, with
$u(t)\in D(H_{\alpha(t)})$ for all $t\in\mathbb{R}$\thinspace.
\end{enumerate}
\end{proposition}

\noindent\textbf{Proof: \ }{} \textbf{1)} The solution of $(\ref{eq.mildint})$
is obtained via the equation (\ref{eq.mildfour2})
\[
\widehat{u}(t,\xi)=e^{-it|\xi|^{2}}\widehat{u}_{0}(\xi)-ie^{-it|\xi|^{2}}%
\mathcal{F}\left[  q1_{[0,t]}\right]  (-|\xi|^{2})\,.
\]
Let us check that we have the required regularity for $u$. Applying Lemma
\ref{le.chgvar} to (\ref{eq.mildfour2}) implies
\begin{equation}
\Vert u(t)\Vert_{H^{1}}\leq\Vert u_{0}\Vert_{H^{1}}+C\Vert q1_{[0,t]}%
\Vert_{H^{\frac{1}{4}}}.\label{cauch2}%
\end{equation}
Lemma \ref{paraproduct}-a) and Lemma \ref{le.10t} yield
\begin{align*}
\Vert q1_{[0,t]}\Vert_{H^{\frac{1}{4}}} &  \lesssim\Vert q\Vert_{H^{\frac
{1}{4}}}\Vert1_{[0,t]}\Vert_{B_{2,\infty}^{\frac{1}{2}}\cap L^{\infty}}\\
&  \leq C_{t}\Vert q\Vert_{H^{\frac{1}{4}}}.
\end{align*}
This proves that $u\in L_{loc}^{\infty}(\mathbb{R};H^{1}).$ It remains to
prove the continuity in time of $u$. First notice that we need for this
purpose to prove only the continuity in time of $v(t):=u(t)-e^{it\Delta}%
u_{0}.$ This will be done in two steps. In the first one we deal with the case
$\alpha\in H_{loc}^{\frac{3}{4}}$\thinspace. In the second one, we go back to
the case $\alpha\in H_{loc}^{\frac{1}{4}}.$\newline$\bullet$ \textit{Case
}$\alpha\in H_{loc}^{\frac{3}{4}}$. Remark that according to Proposition
\ref{sol1}-2) we can construct a unique solution $q\in H_{loc}^{\frac{3}{4}}$
for the problem (\ref{eq.convolsupp}). An easy computation gives for
$t,t^{\prime}\in\mathbb{R}$
\begin{align*}
\Vert v(t)-v(t^{\prime})\Vert_{H^{1}}^{2} &  \lesssim\int_{\mathbb{R}}\sin
^{2}\big((t-t^{\prime2}/2)\big)(1+|\xi|^{2})\big|\mathcal{F}\left[
q1_{[0,t]}\right]  (-|\xi|^{2})\big|^{2}d\xi\\
&  +\Vert q(1_{[0,t]}-1_{[0,t^{\prime}]})\Vert_{H^{\frac{1}{4}}}^{2}.
\end{align*}
Using the fact $\sin^2 x\leq|x|^{\varepsilon},\forall\varepsilon
\in\lbrack0,1],$ and Lemma \ref{le.chgvar} gives
\[
\int_{\mathbb{R}}\sin^{2}\big((t-t^{\prime2}/2)\big)(1+|\xi|^{2}%
)\big|\mathcal{F}\left[  q1_{[0,t]}\right]  (-|\xi|^{2})\big|^{2}d\xi
\lesssim|t-t^{\prime\varepsilon}\Vert q1_{[0,t]}\Vert_{H^{\frac{1}{4}%
+\frac{\varepsilon}{2}}}.
\]
It suffices now to use Lemma \ref{paraproduct}-a)%
\[
\int_{\mathbb{R}}\sin^{2}\big((t-t^{\prime2}/2)\big)(1+|\xi|^{2}%
)\big|\mathcal{F}\left[  q1_{[0,t]}\right]  (-|\xi|^{2})\big|^{2}d\xi
\lesssim|t-t^{\prime\varepsilon}\Vert q\Vert_{H^{\frac{1}{4}+\frac
{\varepsilon}{2}}}.
\]
For the second term we use again Lemma \ref{paraproduct}-c) combined with the
proof of Lemma \ref{le.10t}
\begin{align*}
\Vert q(1_{[0,t]}-1_{[0,t^{\prime}]})\Vert_{H^{\frac{1}{4}}} &  \lesssim\Vert
q\Vert_{H^{\frac{1}{4}+\varepsilon}}\Vert1_{[0,t]}-1_{[0,t^{\prime}]}%
\Vert_{H^{\frac{1}{2}-\varepsilon}}\\
&  \lesssim\Vert q\Vert_{H^{\frac{1}{4}+\varepsilon}}|t-t^{\prime\varepsilon}%
\end{align*}
This concludes the proof of the time continuity of $u$ when $\alpha\in
H_{loc}^{\frac{3}{4}}.$\newline$\bullet$ \textit{{Case }}$\alpha\in
H_{loc}^{\frac{1}{4}}$. We smooth out the function $\alpha$ leading to a
sequence of smooth functions $\alpha_{n}$ that converges strongly to $\alpha$
in $H_{loc}^{\frac{1}{4}}$. To each $\alpha_{n}$ we associate the unique
solutions $q_{n}$ and $u_{n}.$ From the first step $u_{n}$ belongs to
$C(\mathbb{R};H^{1})$. Similarly to (\ref{cauch2}) we get for $n,m\in
\mathbb{N}$
\[
\Vert u_{n}-u_{m}\Vert_{L_{[-T,T]}^{\infty}H^{1}}\leq C_{T}\Vert q_{n}%
-q_{m}\Vert_{H^{\frac{1}{4}}}.
\]
By Proposition \ref{sol1}-a), $\{q_{n}\}$ is a Cauchy sequence in $H^{\frac
{1}{4}}$ and thus $\{u_{n}\}$ converges uniformly to $u$ in $L_{T}^{\infty
}H^{1}$. This gives that $u\in C([-T,T],H^{1}),$ for every $T>0.$%
\newline\textbf{2)} Recall from (\ref{eq.timeder2}) that
\[
i\partial_{t}\widehat{u}(t,\xi)=e^{-it|\xi|^{2}}\mathcal{F}({H_{\alpha(0)}u_{0}%
})(\xi)+e^{-it|\xi|^{2}}\mathcal{F}\left[  (\partial_{s}q)1_{[0,t]}\right]
(-|\xi|^{2})\,.
\]
Since $u_{0}\in D(H_{\alpha(0)})$ then the first term of the r.h.s belongs to
$C(\mathbb{R};L^{2}).$ On the other hand we have $D(H_{\alpha(0})\subset
H^{\frac{3}{2}-d}$ for any $d>0$. It follows from Proposition \ref{propeq}%
-\textbf{1)} that we can construct a unique solution $q\in H^{\frac{3}{4}}$.
Now, let $w(t):=i\partial_{t}u-e^{it\Delta}H_{\alpha(0)}u_{0}.$ Then Lemma \ref{le.chgvar}
yields
\[
\Vert w(t)\Vert_{L^{2}}\lesssim\Vert q^{\prime}\,1_{[0,t]}\Vert_{H^{-\frac
{1}{4}}}.
\]
Lemma \ref{paraproduct}-a) and Lemma \ref{le.10t} imply
\begin{align*}
\Vert q^{\prime}\,1_{[0,t]}\Vert_{H^{-\frac{1}{4}}} &  \lesssim\Vert
q^{\prime}\Vert_{H^{-\frac{1}{4}}}\Vert1_{[0,t]}\Vert_{B_{2,\infty}^{\frac
{1}{2}}\cap L^{\infty}}\\
&  \leq C_{T}\Vert q\Vert_{H^{\frac{3}{4}}}.
\end{align*}
Thus we get for every $t\in\lbrack-T,T]$
\begin{equation}
\Vert w(t)\Vert_{L^{2}}\leq C_{T}\Vert q\Vert_{H^{\frac{3}{4}}}.\label{cauch3}%
\end{equation}
It follows that $w\in L_{loc}^{\infty}(\mathbb{R};L^{2})$. To prove the
continuity in time of $w$ we use the same argument as for the first point of
this proposition. We start with a smooth function $\alpha$, that is $\alpha\in
H_{loc}^{\frac{3}{4}+\varepsilon}$. This gives according to Proposition
\ref{sol1}-3) a unique solution $q\in H^{\frac{3}{4}+\varepsilon}$. We have
the following estimate, obtained similarly to case $\alpha\in H_{loc}%
^{\frac{3}{4}}$ discussed above,
\[
\Vert w(t)-w(t^{\prime})\Vert_{L^{2}}\lesssim|t-t^{\prime}|^\varepsilon\Vert
q^{\prime}1_{[0,t]}\Vert_{H^{-\frac{1}{4}+\frac{\varepsilon}{2}}}+\Vert
q^{\prime}(1_{[0,t]}-1_{[0,t^{\prime}]})\Vert_{H^{-\frac{1}{4}}}.
\]
Using Lemma \ref{paraproduct}-a) with $s=-\frac{1}{4}+\frac{\varepsilon}{2}$ gives
\[
\Vert q^{\prime}1_{[0,t]}\Vert_{H^{-\frac{1}{4}+\frac{\varepsilon}{2}}%
}\lesssim\Vert q\Vert_{H^{\frac{3}{4}+\varepsilon}}.
\]
For the second term of the r.h.s we use Lemma \ref{paraproduct}-c- with
$s=-\frac{1}{4}+\varepsilon,s^{\prime}=\frac{1}{2}-\varepsilon$ and Lemma \ref{le.10t}
\[
\Vert q^{\prime}(1_{[0,t]}-1_{[0,t^{\prime}]})\Vert_{H^{-\frac{1}{4}}}%
\lesssim|t-t^{\prime}|^\varepsilon\Vert q\Vert_{H^{\frac{3}{4}+\varepsilon}}.
\]
This achieves the proof of the continuity of $w$ in time for $\alpha\in
H_{loc}^{\frac{3}{4}+\varepsilon}.$ Now for $\alpha\in H_{loc}^{\frac{3}{4}}$
we do like the first point of the proposition: we smooth out $\alpha$ and we
use the continuity dependence of $q$ with respect to $\alpha$ stated in
Proposition \ref{sol1}-2) combined with the estimate (\ref{cauch3}). By
writing $i\partial_{t}u=H_{\alpha(t)}u(t)$ we get that for every
$t\in\mathbb{R},$ $u(t)\in D(H_{\alpha(t)}).$ {\hfill$\square$\newline}{}
\appendix

\section{Paraproducts and product laws}

\label{se.law} The aim of this section is to prove some product laws used in
the proof of the main results. For this purpose we first recall some basic
ingredients of the paradifferential calculus. Start with the dyadic partition
of the unity: there exists two radial positive functions $\chi\in
\mathcal{D}(\mathbb{R})$ and $\varphi\in\mathcal{D}(\mathbb{R}\backslash
{\{0\}})$ such that
\[
\chi(\xi)+\sum_{q\geq0}\varphi(2^{-q}\xi)=1,\quad\forall\xi\in\mathbb{R}.
\]
For every tempered distribution $v\in{\mathcal{S}}^{\prime}$, set
\[
\Delta_{-1}v=\chi(D)v~;\,\forall q\in\mathbb{N},\;\Delta_{q}v=\varphi
(2^{-q}D)v\quad\text{and}\;S_{q}=\sum_{j=-1}^{q-1}\Delta_{j}.
\]
For more details see for instance \cite{Che}\cite{Bon}. Then Bony's
decomposition of the product $uv$ is given by
\[
uv=T_{u}v+T_{v}u+R(u,v),
\]
with
\[
T_{u}v=\sum_{q}S_{q-1}u\Delta_{q}v\quad\text{and}\quad R(u,v)=\sum
_{|q^{\prime}-q|\leq1}\Delta_{q}u\Delta_{q^{\prime}}v\,.
\]
Let us now recall the definition of Besov spaces through dyadic decomposition.
For $(p,r)\in\lbrack1,+\infty]^{2}$ and $s\in\mathbb{R}$, the space
$B_{p,r}^{s}$ is the set of tempered distribution $u$ such that
\[
\Vert u\Vert_{B_{p,r}^{s}}:=\Big(2^{qs}\Vert\Delta_{q}u\Vert_{L^{p}%
}\Big)_{\ell^{r}}<+\infty.
\]

This definition does not depend on the choice of the dyadic decomposition. One
can further remark that the Sobolev space $H^{s}$ coincides with $B_{2,2}^{s}%
$. Below is the Bernstein lemma that will be used for the proof of product
laws and which is a straightforward application of convolution estimates and
Fourier localization.

\begin{lemme}
\label{lb}\; There exists a constant $C$ such that for $q,k\in\mathbb{N},$
$1\leq a\leq b$ and for $f\in L^{a}(\mathbb{R})$,
\begin{align*}
\sup_{|\alpha|=k}\|\partial^{\alpha}S_{q}f\|_{L^{b}}  &  \leq C^{k}%
\,2^{q(k+\frac{1}{a}-\frac{1}{b})}\|S_{q}f\|_{L^{a}},\\
\ C^{-k}2^{qk}\|{\Delta}_{q}f\|_{L^{a}}  &  \leq\sup_{|\alpha|=k}%
\|\partial^{\alpha}{\Delta}_{q}f\|_{L^{a}}\leq C^{k}2^{qk}\|{\Delta}%
_{q}f\|_{L^{a}}.
\end{align*}

\end{lemme}

The following product laws have been used intensively in the proof of our main result.

\begin{lemme}
\label{paraproduct} In dimension $d=1$ the product $(u,v)\mapsto uv$ is
bilinear continuous

\begin{description}
\item[a)] from $H^{s}\times(B_{2,\infty}^{\frac12}\cap L^{\infty})$ to $H^{s}$
as soon as $|s|<\frac12$\,;

\item[b)] from $H^{s} \times B_{\infty,\infty}^{s+\varepsilon}$ to $H^{s}$ as
soon as $s\geq0$ and $\varepsilon> 0.$

\item[c)] from $H^{s}\times H^{s^{\prime}}$ to $H^{s+s^{\prime}-\frac12}$ as soon as
$s,s^{\prime}<\frac12$ and $s+s^{\prime}>0.$

\item[d)] For $s\geq0$, $H^{s}\cap L^{\infty}$ is an algebra. For $s>\frac12$,
$H^{s}$ is an algebra.
\end{description}
\end{lemme}

\noindent\textbf{Proof: \ } \textbf{a)} Using the definition and Bernstein
lemma we obtain
\begin{align*}
\|T_{u} v\|_{H^{s}}^{2}  &  \lesssim\sum_{q}2^{2qs}\|S_{q-1}u\|_{L^{\infty}%
}^{2}\|\Delta_{q} v\|_{L^{2}}^{2}\\
&  \lesssim\|v\|_{B_{2,\infty}^{\frac12}}^{2}\sum_{q}2^{2q(s-\frac
12)}\|S_{q-1}u\|_{L^{\infty}}^{2}\\
&  \lesssim\|v\|_{B_{2,\infty}^{\frac12}}^{2}\sum_{q}\Big(\sum_{p\le
q-1}2^{q(s-\frac12)}2^{\frac p2}\|\Delta_{p} u\|_{L^{2}}\Big)^{2}\\
&  \lesssim\|v\|_{B_{2,\infty}^{\frac12}}^{2}\sum_{q}\Big(\sum_{p\le
q-1}2^{(q-p)(s-\frac12)}(2^{ps}\|\Delta_{p} u\|_{L^{2}})\Big)^{2}\\
&  \lesssim\|v\|_{B_{2,\infty}^{\frac12}}^{2}\|u\|_{H^{s}}^{2}.
\end{align*}
We have used in the last line the convolution law $\ell^{1}\star\ell^{2}%
\to\ell^{2}.$\newline For the second term $T_{v} u$ we use the fact that
$S_{q-1}$ maps $L^{\infty}$ to itself uniformly with respect to $q$.
\begin{align*}
\|T_{v} u\|_{H^{s}}^{2}  &  \lesssim\sum_{q}2^{2qs}\|S_{q-1}v\|_{L^{\infty}%
}^{2}\|\Delta_{q} u\|_{L^{2}}^{2}\\
&  \lesssim\|v\|_{L^{\infty}}^{2}\|u\|_{H^{s}}^{2}.
\end{align*}
To estimate the remainder term we use the fact
\[
\Delta_{q}\mathcal{R}(u,v)=\sum_{%
\genfrac{}{}{0pt}{}{j\geq q-4}{|j-j^{\prime}|\le1}%
}\Delta_{q}(\Delta_{j}u\Delta_{j^{\prime}}v).
\]
According to Bernstein lemma one gets
\begin{align*}
2^{qs}\|\Delta_{q}(\mathcal{R}(u,v))\|_{L^{2}}  &  \lesssim2^{q(s+\frac
12)}\sum_{%
\genfrac{}{}{0pt}{}{j\geq q-4}{|j-j^{\prime}|\le1}%
}\|\Delta_{j}u\|_{L^{2}}\|\Delta_{j^{\prime}}v\|_{L^{2}}\\
&  \lesssim\sum_{%
\genfrac{}{}{0pt}{}{j\geq q-4}{|j-j^{\prime}|\le1}%
}2^{(q-j)(s+\frac12)}2^{js}\|\Delta_{j}u\|_{L^{2}}2^{j^{\prime}\frac
12}\|\Delta_{j^{\prime}}v\|_{L^{2}}\\
&  \lesssim\|v\|_{B_{2,\infty}^{\frac12}}\sum_{j\geq q-4}2^{(q-j)(s+\frac
12)}2^{js}\|\Delta_{j}u\|_{L^{2}}.
\end{align*}
It suffices now to apply the convolution inequalities.

\textbf{b)} First remark that the case $s=0$ is obvious: $L^{2}\times
L^{\infty}\to L^{2}$ and $B_{\infty,\infty}^{\varepsilon}\hookrightarrow
L^{\infty}.$ Hereafter we consider $s>0$\,. To estimate the first paraproduct,
use the embedding $B_{\infty,\infty}^{s+\varepsilon}\hookrightarrow
B_{\infty,2}^{s},$ for $\varepsilon>0.$
\begin{align*}
\|T_{u} v\|_{H^{s}}^{2}  &  \lesssim\sum_{q}2^{2qs}\|S_{q-1}u\|_{L^{2}}%
^{2}\|\Delta_{q} v\|_{L^{\infty}}^{2}\\
&  \lesssim\|u\|_{L^{2}}\|v\|_{B_{\infty,2}^{s}}\\
&  \lesssim\|u\|_{L^{2}}\|v\|_{B_{\infty,\infty}^{s+\epsilon}}%
\end{align*}
For the second term we use the result obtained in the part $a)$:
\begin{align*}
\|T_{v} u\|_{H^{s}}^{2}  &  \lesssim\sum_{q}2^{2qs}\|S_{q-1}v\|_{L^{\infty}%
}^{2}\|\Delta_{q} u\|_{L^{2}}^{2}\\
&  \lesssim\|v\|_{L^{\infty}}^{2}\|u\|_{H^{s}}^{2}\\
&  \lesssim\|v\|_{B_{\infty,\infty}^{s+\epsilon}}^{2}\|u\|_{H^{s}}^{2}%
\end{align*}
To estimate the remainder term we write
\begin{align*}
2^{qs}\|\Delta_{q}(\mathcal{R}(u,v))\|_{L^{2}}  &  \lesssim2^{qs}\sum_{%
\genfrac{}{}{0pt}{}{j\geq q-4}{|j-j^{\prime}|\le1}%
}\|\Delta_{j}u\|_{L^{2}}\|\Delta_{j^{\prime}}v\|_{L^{\infty}}\\
&  \lesssim\|v\|_{L^{\infty}}\sum_{%
\genfrac{}{}{0pt}{}{j\geq q-4}{|j-j^{\prime}|\le1}%
}2^{(q-j)s}2^{js}\|\Delta_{j}u\|_{L^{2}}.
\end{align*}
Since $s>0$ then we obtain by using the convolution inequalities
\[
\|\mathcal{R}(u,v)\|_{H^{s}}\le\|v\|_{L^{\infty}}\|u\|_{H^{s}}.
\]
\textbf{c), d)} These results are standard, see for example \cite{Che}.

{\hfill$\square$\newline}

\section{Sobolev and Besov regularity of cut-offs}

\label{se.sobcut}

\begin{lemme}
\label{le.10t}

\begin{enumerate}
\item For any $\nu<1/2$, $t\mapsto1_{[0,t]}(s)$ belongs to $C(\mathbb{R}%
_{+};H^{\nu}(\mathbb{R}))$\thinspace. More precisely we have for $|t-t^{\prime}|\le 1$
$$
\|1_{[0,t]}-1_{[0,t^{\prime}]}\|_{H^{\nu}}\lesssim |t-t^{\prime}|^{\frac12-\nu}.
$$

\item The map $t\mapsto1_{[0,t]}(s)$ belongs to $L^{\infty}(\mathbb{R}%
_{+};B_{2,\infty}^{\frac12}(\mathbb{R}))$. \end{enumerate}
\end{lemme}

\noindent\textbf{Proof: \ }{} \textbf{1)} The Fourier transform of
$1_{[0,t]}(s)$ equals $\mathcal{F}(1_{[0,t]})(\tau)=\frac{e^{-i\tau t}%
-1}{-i\tau}$. One gets for $\nu\in\lbrack0,1/2)$
\[
\int_{\mathbb{R}}(1+\tau^{2})^{\nu}\left\vert \mathcal{F}(1_{[0,t_{1}%
]})-\mathcal{F}(1_{[0,t_{2}]})\right\vert ^{2}(\tau)~d\tau=4\int_{\mathbb{R}%
}(1+\tau^{2})^{\nu}\frac{\left\vert \sin(\tau|t_{2}-t_{1}|/2)\right\vert ^{2}%
}{\tau^{2}}~d\tau:=I.
\]
Let $\lambda>1$ then
\begin{align*}
I &  \lesssim|t_{2}-t_{1}|^{2}\int_{0}^{\lambda}\tau^{2\nu}d\tau+\int
_{\lambda}^{+\infty}\tau^{2\nu-2}d\tau\\
&  \lesssim|t_{2}-t_{1}|^{2}\lambda^{2\nu+1}+\lambda^{2\nu-1}.
\end{align*}
Choosing judiciously $\lambda$ then we obtain for $|t_{2}-t_{1}|\leq1$%

\[
\left\|  1_{[0,t_{1}]}-1_{[0,t_{2}]}\right\|  _{H^{\nu}}\leq C_{\nu}\left|
t_{2}-t_{1}\right|  ^{1/2-\nu}\,.
\]

\textbf{2)} We set $f_{t}(s):=1_{[0,t]}(s)$, then have
\begin{align*}
\|f_{t}\|_{B_{2,\infty}^{\frac12}}^{2}  &  \le\|f\|_{L^{2}}^{2}+\max
_{q\in\mathbb{N}}2^{q}\int_{2^{q}\le|\tau|\le2^{q+1}}|\widehat{f_{t}}%
(\tau)|^{2}d\tau\\
&  \lesssim|t|+\max_{q\in\mathbb{N}}2^{q}\int_{2^{q}\le|\tau|\le2^{q+1}}%
|\tau|^{-2}d\tau\\
&  \lesssim|t|+1.
\end{align*}

{\hfill$\square$\newline}{}


\begin{thebibliography}{99}                                                                                               %


\bibitem {Adami}R.~Adami and A.~Teta, \emph{A class of nonlinear Schr\"{o}dinger
Equations with concentrated nonlinearitie}. J. Func. Anal. \textbf{180},
148-175, 2001.

\bibitem {Albeverio}S.~Albeverio, F.~Gesztesy, R.~H\"{o}gh-Krohn and
H.~Holden. \emph{Solvable Models in Quantum Mechanics. 2nd ed. with an
appendix by P. Exner}. AMS, Providence R.I, 2005.

\bibitem {Bon}J.-M.~Bony, \emph{Calcul symbolique et propagation des
singularit{\'e}s pour les {\'e}quations aux d{\'e}riv{\'e}es partielles non
lin{\'e}aires}.  \newblock Ann. Sci. {\'E}cole
Norm. Sup. (4), \textbf{14}, no. 2, 209--246, 1981.

\bibitem {Che}J.-Y.~Chemin, \emph{Perfect incompressible fluids. Translated
from the 1995 French original by Isabelle Gallagher and Dragos Iftimie}.
Oxford Lecture Series in Mathematics and its Applications, 14. The Clarendon
Press, Oxford University Press, New York, 1998.

\bibitem {Fattorini}H.O.~Fattorini, A.~Kerber. \emph{The Cauchy problem}.
Cambridge University Press, 1984.

\bibitem {Kat}T.~Kato. \emph{Integration of the equation of evolution in a Banach
spac}, J. Math. Soc. Japan \textbf{5}, 208-234, 1953.

\bibitem {Kato1}T.~Kato, \emph{Linear evolution equations of 'hyperbolic' type}. J.
Fac. Sci. Univ. Tokyo Sect. I, A Math \textbf{17}, 241-258, 1970.

\bibitem {Kys}J.~Kisy\'{n}ski, \emph{Sur les op\'{e}rateurs de Green des
probl\`{e}mes de Cauchy abstraits}. Studia Mathematica \textbf{23}, 285-328, 1964.

\bibitem {Reed}M.~Reed and B.~Simon. \emph{Methods of Modern Mathematical
Physics}, Vol. 2. Academic Press, New York, 1975.

\bibitem {Yaf3}M.~Sayapova and D.~Yafaev \emph{The evolution operator for time
dependent potentials of zero radius}. Trudy Mat. Inst. Steklov. 159, 167--174, 1983.

\bibitem {Simon}B.~Simon. \emph{Quantum Machanics for Hamiltonians defined as
quadratic forms}. Princeton University Press, Princeton, New Jersey, 1971.

\bibitem {Yaf1}D.~Yafaev, \emph{On \textquotedblleft
eigenfunctions\textquotedblright\ of a time-dependent Schr\"{o}dinger
equation}. (Russian), Teoret. Mat. Fiz. \textbf{43}, no. 2, 228--239, 1980.

\bibitem {Yaf2}D.~Yafaev, \emph{Scattering theory for time-dependent zero-range
potentials}. Ann.~I.H.P. Physique Th\'{e}orique, \textbf{40}, 1984.
\end{thebibliography}
\end{document}